\documentclass[showpacs,amsmath,amsfonts,amssymb,aps,superscriptaddress]{revtex4}

\usepackage{makecell, tabularx}
\usepackage[dvipsnames]{xcolor}
\usepackage{comment}
\usepackage{enumitem}
\usepackage{graphicx}% Include figure files
\usepackage{float}
\usepackage{dcolumn}% Align table columns on the decimal point
\usepackage{bm}% bold math
\usepackage{mathrsfs}
\usepackage{amsmath}

\usepackage{amsfonts}
\usepackage{dsfont}

\begin{document}

\title{A Unified Spectral Approach for Quasinormal Modes of Morris-Thorne Wormholes}
\author{Davide Batic}
\email{davide.batic@ku.ac.ae}
\affiliation{
Mathematics Department, Khalifa University of Science and Technology, PO Box 127788, Abu Dhabi, United Arab Emirates}
\author{Denys Dutykh}
\email{denys.dutykh@ku.ac.ae}
\affiliation{
Mathematics Department, Khalifa University of Science and Technology, PO Box 127788, Abu Dhabi, United Arab Emirates}
\affiliation{Causal Dynamics Pty Ltd, Perth, Australia}

\date{\today}

\begin{abstract}
In this paper, we undertake a comprehensive examination of quasinormal modes (QNMs) linked to Morris-Thorne, also known as Bronnikov-Ellis wormholes, delving into scalar, electromagnetic, and gravitational perturbations using the spectral method. Our research corrects inaccuracies previously reported in the literature and addresses areas where the Wentzel--Kramers--Brillouin (WKB) approximation proves inadequate. Moreover, we introduce and evaluate a novel spectral technique designed to consolidate recent advancements in formulating QNM boundary conditions at both the wormhole throat and space-like infinity. This innovative approach bridges critical gaps in existing methodologies and enhances the accuracy and applicability of QNM analysis in the study of wormhole physics.
\end{abstract}
%\keywords{Suggested keywords}% Use the 'showkeys' class option if the keyword
\maketitle

\section{Introduction}

QNMs serve as a fundamental concept in gravitational wave analysis, offering a window into the dynamics of dissipative systems within our universe. In any system where vibrations occur, energy decay is an inevitable phenomenon. This dissipation is characterised by complex, discrete, characteristic oscillation patterns, known as QNMs, which are defined under specific outgoing boundary conditions. The significance of QNMs becomes especially pronounced in the context of gravitational wave observations. Extracting these modes from observed signals allows us to assess the stability of spacetime through the temporal decay of these signals. Furthermore, QNMs are uniquely determined by the intrinsic parameters of their source, such as mass, charge, and angular momentum, making them a tool for probing the fundamental properties of astrophysical objects such as black holes. They also offer profound insights into the governing theories of gravity by analysing the spacetime metric involved.  Our present work focuses on the QNMs arising from scalar, electromagnetic, and gravitational perturbations in the presence of Morris-Thorne wormholes.

Wormholes are intriguing hypothetical shortcuts through spacetime that may have profound implications for space travel and our understanding of the universe's fabric. They are also called Einstein-Rosen bridges, even though Flamm first discovered them in 1916 \cite{Flamm1916PZ} and later rediscovered by Einstein and Rosen in 1935 \cite{Einstein1935PR} in their attempt to represent particles as bridges linking two identical spatial sheets. In 1962, \cite{Fuller1962PR} demonstrated that such wormholes are unstable within the same universe, pinching off too quickly to allow light from one region to reach another. Yet, the narrative began to shift as \cite{Ellis1973JMP, Ellis1974JMP, Bronnikov1973APP, Ellis1979GRG} introduced the notion of traversable wormholes, facilitated by the inclusion of exotic matter or unconventional scalar fields. These investigations unveiled that phantom fields, i.e. scalar entities with negative energy densities, could pave the way for wormholes that could be traversed. The discussion broadened with alternative gravitational theories like Einstein-scalar-Gauss-Bonnet gravity \cite{Kanti2011PRL, Kanti2012PRD, Antoniou2020PRD}, suggesting wormholes could exist without exotic matter. These theories indicated that violating energy conditions, essential for wormhole creation, could be achieved through gravitational interactions alone, supported by studies on Dirac particles and 3-form fields \cite{BlazquezSalcedo2020EPJC, BlazquezSalcedo2021PRL, BlazquezSalcedo2022EPJC, Konoplya2022PRL, Barros2018PRD, BouhmadiLopez2021JCAP}.

The goal of detecting wormholes has sparked intense scholarly activity, transcending their theoretical foundations. Detection strategies encompass a spectrum of approaches, from observing gravitational lensing effects \cite{Abe2010AJ, Bambi2013PRD, Batic2015PRD, Cramer1995PRD, Kuhfittig2014EPJC, Nakajima2012PRD, Nandi2006PRD, Lukmanova2018PRD, Perlick2004PRD, Safonova2002PRD, Takahashi2013AJL, Toki2011AJ, Tsukamoto2012PRD, Tsukamoto2017PRD, Tsukamoto2016PRD, Tsukamoto2017PRD95, Tsukamoto2022EPJC}  and the distinctive shadows cast by wormholes \cite{Bambi2013PRD, BouhmadiLopez2021JCAP, Guerrero2022PRD, Gyulchev2018EPJC, Nedkova2013PRD, Ohgami2015PRD, Shaikh2018PRD} to analyzing the radiation from accretion disks \cite{Bambi2013PRDa, Deligianni2021PRD, Deligianni2021PRDa, Harko2008PRD, Harko2008PRDa, Lamy2018CQG, Zhou2016PRD}. Notably, the interaction of scalar fields with wormholes, manifesting in unique transmission, reflection, and absorption characteristics, has emerged as a compelling method to differentiate wormholes from black holes, providing discernible markers \cite{Azad2020EPJC}. 

Furthermore, it is worth mentioning that a novel analytical approach via the Heun equation and numerical techniques for the analysis of the boundary problems corresponding to QNMs of black holes and other simple models of compact objects has been developed by \cite{Fiziev2006CQG}. Last but not least, \cite{Konoplya2018PLB} has shown how QNMs can be used to discern the shape of a wormhole, providing a robust framework for analyzing their stability and observational signatures.

The groundbreaking discovery of gravitational waves by the LIGO-Virgo collaborations \cite{Abbott2016PRL} has revolutionized the study of compact astrophysical objects, offering an innovative method to potentially observe wormholes via their QNMs \cite{Berti2009CQG, Kokkotas1999LR, Konoplya2011RMP}, which not only encode essential information about the objects' physical characteristics but also provides a distinctive spectral signature for black holes, neutron stars, and possibly wormholes. Research on the QNMs of wormholes has been extensively undertaken across diverse studies, highlighting their significance in theoretical physics \cite{Aneesh2018PRD, BlazquezSalcedo2018PRD, Churilova2020PLB, Gonzalez2022PRD, Jusufi2021GRG, Kim2008PTPS, Konoplya2005PRD, Konoplya2010PRD, Konoplya2016JCAP, Konoplya2018PLB, Konoplya2019PRD, Voelkel2018CQG}. An intriguing dimension of this research involves the inverse problem, where the objective is to infer the geometric structure of a wormhole based on the characteristics of its QNMs \cite{Konoplya2018PLB, Voelkel2018CQG}. This contrasts with black holes, where different effective potentials can yield identical QNM spectra, complicating their analysis \cite{Chandrasekhar1998}. QNMs have proven instrumental in detailed investigations of wormhole geometries, particularly for symmetric Ellis-Bronnikov wormholes. In these cases, QNMs facilitate the reconstruction of spacetime metrics near the wormhole's throat, offering concrete examples of their utility \cite{Konoplya2018PLB}. Several studies \cite{Azad2023PRD, BlazquezSalcedo2018PRD, Kim2008PTPS, Konoplya2005PRD} have explored the QNMs of the Ellis-Bronnikov wormhole family, including analyses on scalar, axial, and radial perturbations, performed using the WKB approximation or by numerical methods using a collocation method for systems of ordinary differential equations with error estimation and adaptive mesh selection \cite{Ascher1979MC}. These studies underscore the potential of QNMs as a diagnostic tool for understanding the properties and structures of wormholes and differentiating between various compact objects in the cosmos. 

Our study marks a significant advancement in the vast literature on QNMs, particularly in analyzing Morris-Thorne wormholes across scalar, electromagnetic, and gravitational perturbations. We not only correct previous analytical inaccuracies but also explore areas beyond the reach of the WKB approximation. Additionally, we introduce a novel spectral method that optimizes the formulation of QNM boundary conditions, offering a comprehensive approach that surpasses the methodology presented in \cite{Gonzalez2022PRD}. This work presents a unified spectral method for computing the QNMs of Morris-Thorne wormhole perturbations, enhancing both precision and efficiency. Lastly, our numerical method has been successfully validated by reproducing previously established results for QNMs of Schwarzschild black holes and black holes inspired by noncommutative geometry \cite{Batic2024EPJC}.

The structure of our paper is as follows: Section II introduces the Morris-Thorne wormhole metric and its effective potential for various perturbations. Section III is devoted to presenting our unified spectral method, optimizing and extending the analysis beyond \cite{Gonzalez2022PRD}'s approach. Section IV details the numerical methods underpinning our analysis, setting the stage for the results presented in Section V. Here, we validate our methodology against \cite{Gonzalez2022PRD}, correcting and expanding upon the findings in \cite{Kim2008PTPS, Gonzalez2022PRD}. The conclusion in Section V reflects on our contributions to the field and proposes future research directions.

\section{THE MORRIS-THORNE WORMHOLE: Metric and Equations of motion}

In this study, we investigate the behaviour of a massless scalar field denoted as $\psi$ within the spacetime of a Morris-Thorne wormhole. The spacetime is described by a metric, expressed in natural units where $c = G_N = 1$, given by the following line element \cite{Morris1988AJP}
\begin{equation}\label{metric}
  ds^2 = -e^{2\Lambda(r)}dt^2+\frac{dr^2}{1-\frac{b(r)}{r}} + r^2d\vartheta^2 + r^2\sin^2{\vartheta}d\varphi^2, \quad \vartheta\in[0,\pi], \quad \varphi\in[0,2\pi).
\end{equation}
Here, $\Lambda(r)$ and $b(r)$ are defined as the lapse/redshift and shape functions, respectively. These functions play crucial roles in the geometry of wormhole spacetime. Our analysis primarily focuses on cases where $\Lambda(r) = 0$ and $b(r)$ follow the specific form $b(r) = b_0^2/r$, where $b_0$ denotes the wormhole throat's radius. Note that this scenario corresponds precisely to the Ellis-Bronnikov wormhole \cite{Ellis1973JMP,Ellis1974JMP,Ellis1979GRG,Bronnikov1973APP}, which predates the Morris-Thorne wormhole by several years. Within the coordinate system $(t, r, \theta, \phi)$, the radial coordinate $r$ must meet the requirement $r > b_0$. Extending $r$ into the range $0 < r < b_0$ is not feasible, as the metric tensor undergoes a signature change in this region. However, it might still be possible to find a second pairwise compatible chart, i.e., a chart that overlaps with the previous one over an open subset of the manifold containing the wormhole throat, and such that the transition function between the two charts is a smooth bijection, thereby allowing for the description of the region of the manifold beyond the throat. This aspect will be addressed later on. A study by \cite{Kim2008PTPS} has revealed that within the spacetime framework described by the metric (\ref{metric}), and for the chart $(t,r,\vartheta,\varphi)$ where $r>b_0$, the governing equation for a massless Klein-Gordon field can be expressed in a specific manner. This field is assumed to exhibit a time dependence as $e^{-i\omega t}$ and an angular component represented by spherical harmonics. For $\ell = 0, 1, 2, \ldots$ and various types of perturbations ($s = 0$ for massless scalar perturbation, $s = 1$ for electromagnetic perturbation, and $s = 2$ for vector-type gravitational perturbation), the equation is given by
\begin{equation}\label{ODE01}
  f(r)\frac{d}{dr}\left(f(r)\frac{d\psi_{\omega\ell s}}{dr}\right)+\left[\omega^2-U_{s,\ell}(r)\right]\psi_{\omega\ell s}(r)=0.
\end{equation}
Here, the function $f(r)$ is defined as $f(r) = e^{\Lambda(r)}\sqrt{1-b(r)/r}$, where the prime indicates differentiation with respect to the radial variable. The effective potential $U_{s,\ell}(r)$ takes the general form
\begin{equation}
  U_{s,\ell}(r)=\ell(\ell+1)\frac{e^{2\Lambda(r)}}{r^2}+(1-s)e^{2\Lambda(r)}\left[\frac{1}{r}\left(1-\frac{b(r)}{r}\right)\Lambda^{'}(r)+\frac{(1+2s)b(r)-rb^{'}(r)}{2r^3}\right].
\end{equation}
When the redshift function vanishes, and $b(r)$ takes the form $b(r) = b_0^2/r$, the differential equation (\ref{ODE01}) and the effective potential simplify as follows
\begin{equation}
  \sqrt{1-\frac{b_0^2}{r^2}}\frac{d}{dr}\left(\sqrt{1-\frac{b_0^2}{r^2}}\frac{d\psi_{\omega\ell\epsilon}}{dr}\right)+\left[\omega^2-U_{\epsilon,\ell}(r)\right]\psi_{\omega\ell\epsilon}(r)=0,\quad
U_{\epsilon,\ell}(r)=\frac{\ell(\ell+1)}{r^2}+\epsilon\frac{b_0^2}{r^4},\quad \epsilon = 1-s^2,
\end{equation}
where $r > b_0$. At this point, introducing the rescaling $z = r/b_0$ leads to the following differential equation
\begin{eqnarray}\label{ODEor}
  &&\frac{\sqrt{z^2-1}}{z}\frac{d}{dz}\left(\frac{\sqrt{z^2-1}}{z}\frac{d\psi_{\Omega\ell\epsilon}}{dz}\right)+\left[\Omega^2-V_{\epsilon,\ell}(z)\right]\psi_{\Omega\ell\epsilon}(z)=0,\quad\Omega=\omega b_0\\
  &&V_{\epsilon,\ell}(z)=\frac{\ell(\ell+1)}{z^2}+\frac{\epsilon}{z^4},\quad z>1.
\end{eqnarray}
It is worth noting that equation \eqref{ODEor} can be solved in terms of confluent Heun equations \cite{Ronveaux1995}. The novel technique developed by \cite{Fiziev2006CQG} provides a framework for obtaining QNMs related to such equations. However, we leave the exploration of this approach and its implications for QNMs computations as a future endeavor. The forthcoming sections aim to demonstrate the application of the Spectral Method for extracting QNMs from the Morris-Thorne wormhole, highlighting its efficiency compared to previously utilised techniques in related literature. For example, the study \cite{Kim2008PTPS} employed the WKB approximation to the third order, necessitating a shift to tortoise coordinates $z_*$ within that context. However, a thorough examination of \cite{Kim2008PTPS}'s findings reveals their unreliability. Consequently, our approach is twofold: initially, we employ the spectral method to the problem as formulated in tortoise coordinates, as done in \cite{Kim2008PTPS}, enabling us to uncover QNMs that \cite{Kim2008PTPS} failed to report. Subsequently, we implement the WKB approximation up to the sixth order on the same issue, serving as an independent benchmark to validate our spectral method results and to affirm the unreliability of the numerical outcomes from \cite{Kim2008PTPS}. It's worth noting that we identified several inaccuracies in the formula for the third-order WKB correction; we encourage readers to compare the formulae in \cite{Iyer1987PRD}, \cite{Kim2008PTPS}, and the second expression in (44) by \cite{BlazquezSalcedo2018PRD}. Ultimately, we demonstrate that the Spectral Method's efficacy allows for the extraction of QNMs from an equivalent problem defined over an interval that stretches from the wormhole throat to spatial infinity, circumventing the need for tortoise coordinates.

\section{A Unified Spectral Approach to Quasinormal Modes Across the Morris-Thorne Wormhole}\label{wholeR}

In order to formulate the QNM boundary conditions, a method similar to that employed in \cite{Morris1988AJP, Kim2008PTPS} can be adopted. This involves introducing a new radial coordinate, denoted as $x \in (-\infty,+\infty)$ such that $z = \sqrt{1+x^2}$. By doing so, one can effectively track the propagation of the massless field from positive spacelike infinity to negative spacelike infinity. The resulting differential equation takes the form
\begin{equation}\label{ourODE}
  \frac{d^2\psi_{\Omega\ell\epsilon}}{dx^2} + \left[\Omega^2 -V_\epsilon(x)\right] \psi_{\Omega\ell\epsilon}(x) = 0,
\end{equation}
where the effective potential $V_\epsilon(x)$ is given by
\begin{equation}\label{Veps}
  V_\epsilon(x) = \frac{\ell(\ell+1)}{1+x^2}+\frac{\epsilon}{(1+x^2)^2}.
\end{equation}
In our subsequent analysis, we focus on the computation of the QNMs for the spectral problem described in (\ref{ourODE}). To do so, we represent $\Omega$ as $\Omega = \Omega_R + i\Omega_I$, where $\Omega_I < 0$ ensures the damping of perturbations over time. The appropriate QNM boundary conditions should be such that the radial field exhibits inward radiation in the lower copy of the universe as $x \to -\infty$ and outward radiation in the upper copy of the universe as $x \to +\infty$. The asymptotic behaviour of the solutions to equation \eqref{ourODE} can be deduced using the method outlined in \cite{Olver1994MAA}. For this purpose, we start by observing that \eqref{ourODE} can be expressed as
\begin{equation}
 \frac{d^2\psi_{\Omega\ell\epsilon}}{dx^2} +P(x)\frac{d\psi_{\Omega\ell\epsilon}}{dx}+ \left[\Omega^2 -V_\epsilon(x)\right] \psi_{\Omega\ell\epsilon}(x) = 0
\end{equation}
\begin{equation}
    P(x) = \sum_{\kappa=0}^\infty\frac{\mathfrak{f}_\kappa}{x^k} =0, \qquad
    Q(x) = \sum_{\kappa=0}^\infty\frac{\mathfrak{g}_\kappa}{x^k}=\Omega^2 -V_\epsilon(x)=\mathcal{O}\left(\frac{1}{x^2}\right).
\end{equation}
Given that at least one of the coefficients $\mathfrak{f}_0$, $\mathfrak{g}_0$, $\mathfrak{g}_1$ is nonzero, a formal solution to (\ref{ourODE}) is represented by \cite{Olver1994MAA}
\begin{equation}
    \psi^{(j)}_{\Omega\ell\epsilon}(x) = x^{\mu_j}e^{\lambda_j x}\sum_{\kappa=0}^\infty\frac{a_{\kappa,j}}{x^\kappa}, \qquad j \in \{1,2\},
\end{equation}
where $\lambda_1$, $\lambda_2$, $\mu_1$ and $\mu_2$ are the roots of the characteristic equations
\begin{equation}
   \lambda^2+\mathfrak{f}_0\lambda+\mathfrak{g}_0=0,\quad
   \mu_j=-\frac{\mathfrak{f}_1\lambda_j+\mathfrak{g}_1}{\mathfrak{f}_0+2\lambda_j}.
\end{equation}
A straightforward computation shows that $\lambda_\pm = \pm i\Omega$ and $\mu_\pm =0$.  As a result, the solutions of the radial equation behave as asymptotic plane waves in the form of $e^{\pm i\Omega x}$.
We ensure that the boundary conditions for the QNMs are met by requiring that $\psi_{\Omega\ell\epsilon} \to e^{-i\Omega x}$ as $x \to -\infty$ and $\psi_{\Omega\ell\epsilon} \to e^{i\Omega x}$ as $x \to +\infty$. Based on the analysis above, the remaining part of the radial functions naturally exhibits regular behaviour as \( x \to \pm\infty \). Additionally, it maintains regularity across the entire real axis, as \eqref{ourODE} lacks finite singularities. In a previous study \cite{Kim2008PTPS}, the WKB method was employed, extended to the third order beyond the eikonal approximation, to compute the QNMs. In this work, we instead make use of a Chebyshev-type spectral method \cite{Trefethen2000, Boyd2000}. This technique allows us to verify and expand upon the results obtained in \cite{Kim2008PTPS}. We can proceed in a manner similar to the approach in \cite{Leaver1985PRSLA}, that is, we transform the radial function $\psi_{\Omega\ell\epsilon}(x)$ into a new radial function $\Phi_{\Omega\ell\epsilon}(x)$ such that the QNM boundary conditions described above are automatically implemented and  $\Phi_{\Omega\ell\epsilon}(x)$ has a regular behaviour as $x \to  \pm\infty$. To achieve this, we introduce the following transformation
\begin{equation}\label{Ansatz1}
  \psi_{\Omega\ell\epsilon}(x) = e^{i\frac{2\Omega x}{\pi}\arctan{x}}\Phi_{\Omega\ell\epsilon}(x).
\end{equation}
If we substitute (\ref{Ansatz1}) into (\ref{ourODE}), we end up with the following ordinary differential equation for the radial eigenfunctions, namely
\begin{equation}\label{ODEznone}
    P_2(x)\frac{d^2\Phi_{\Omega\ell\epsilon}}{dx^2} + P_1(x)\frac{d\Phi_{\Omega\ell\epsilon}}{dx} + P_0(x)\Phi_{\Omega\ell\epsilon}(x) = 0
\end{equation}
with
\begin{eqnarray*}
    P_2(x)&=&1,\quad
    P_1(x)=\frac{4i\Omega}{\pi}\left(\arctan{x}+\frac{x}{1+x^2}\right),\quad
    P_0(x)=\Omega^2\mathfrak{Q}(x)+i\Omega\mathfrak{L}(x)-V_\epsilon(x),\\
    \mathfrak{Q}(x)&=&1-\frac{4\left[(1+x^2)\arctan{x}+x\right]^2}{\pi^2(1+x^2)^2},\quad
    \mathfrak{L}(x)=\frac{4}{\pi(1+x^2)^2}.
\end{eqnarray*}
However, our method requires that we transform the differential equation \eqref{ODEznone} onto the finite interval $(-1, 1)$. This is accomplished through a homeomorphic mapping of the real line into the mentioned interval. To achieve this, we introduce a coordinate transformation given by
\begin{equation}\label{trafyx}
  y = \frac{2}{\pi}\arctan{x},
\end{equation}
which maps $\pm\infty$ to $\pm 1$. Hence, \eqref{ODEznone} becomes
\begin{equation}\label{ODEynone}
  S_2(y)\ddot{\Phi}_{\Omega\ell\epsilon}(y) + S_1(y)\dot{\Phi}_{\Omega\ell\epsilon}(y) + S_0(y)\Phi_{\Omega\ell\epsilon}(y) = 0,
\end{equation}
where
\begin{eqnarray}
  S_2(y) &=& \frac{4}{\pi^2}\cos^4{\left(\frac{\pi y}{2}\right)}, \label{S2onone} \\
  S_1(y) &=& \frac{2}{\pi}\cos^2{\left(\frac{\pi y}{2}\right)}\left[\frac{2i\Omega}{\pi}\left(\pi y+\sin(\pi y)\right)-\sin(\pi y)\right],\label{S1onone}\\
  S_0(y) &=& \Omega^2\Sigma_2(y)+i\Omega\Sigma_1(y)+\Sigma_0(y) \label{S0onone}
\end{eqnarray}
with
\begin{eqnarray}
    \Sigma_2(y) &=& 1-\left[y+\frac{1}{\pi}\sin{(\pi y)}\right]^2,\quad
    \Sigma_1(y) =\frac{4}{\pi}\cos^4{\left(\frac{\pi y}{2}\right)},\\
    \Sigma_0(y) &=& -\cos^2{\left(\frac{\pi y}{2}\right)}\left[\ell(\ell+1)+\epsilon\cos^2{\left(\frac{\pi y}{2}\right)}\right].
\end{eqnarray}
We must also require that $\Phi_{\Omega\ell\epsilon}(y)$ is regular at $y = \pm 1$.
\begin{table}%[ht]
\caption{Classification of the points $y = \pm 1$ for the relevant functions defined by  (\ref{S2onone}), (\ref{S1onone}) and (\ref{S0onone}). The abbreviation $z$ ord $n$ stands for zero of order $n$.}
\begin{center}
\begin{tabular}{ | c | c | c | c | c | c | }
\hline
$y$      & $S_2(y)$              & $S_1(y)$       & $S_0(y)$\\ \hline
$-1$     & z \mbox{ord} 4        & z \mbox{ord} 2 & z \mbox{ord} 2 \\ \hline
$+1$     & z \mbox{ord} 4        & z \mbox{ord} 2 & z \mbox{ord} 2\\ \hline
\end{tabular}
\label{tableEinsnone}
\end{center}
\end{table}
Table~\ref{tableEinsnone} shows that the coefficients of the differential equation (\ref{ODEynone}) share a common zero of order $2$ at $y =\pm1$. Hence, in order to apply the spectral method, we need to divide (\ref{ODEynone}) by $(1-y^2)^2$. As a result, we end up with the following differential equation
\begin{equation}\label{ODEhynone}
  M_2(y)\ddot{\Phi}_{\Omega\ell\epsilon}(y) + M_1(y)\dot{\Phi}_{\Omega\ell\epsilon}(y) + M_0(y)\Phi_{\Omega\ell\epsilon}(y) = 0,
\end{equation}
where
\begin{equation}\label{S210honone}
  M_2(y) =  \frac{4\cos^4{\left(\frac{\pi y}{2}\right)}}{\pi^2(1-y^2)^2}, \qquad
  M_1(y) = i\Omega N_1(y)+N_0(y), \qquad
  M_0(y) = \Omega^2 C_2(y)+i\Omega C_1(y)+C_0(y)
\end{equation}
with
\begin{eqnarray}
    N_1(y) &=&  \frac{4\cos^2{\left(\frac{\pi y}{2}\right)}\left[\pi y+\sin{(\pi y)}\right]}{\pi^2(1-y^2)^2}, \quad
    N_0(y) = -\frac{2\sin{(\pi y)}\cos^2{\left(\frac{\pi y}{2}\right)}}{\pi(1-y^2)^2},\label{N0}\\
    C_2(y) &=&\frac{1-\left[y+\frac{1}{\pi}\sin{(\pi y)}\right]^2}{(1-y^2)^2},\label{C2}\\
    C_1(y) &=&\frac{4\cos^4{\left(\frac{\pi y}{2}\right)}}{\pi(1-y^2)^2},\label{C1}\\
    C_0(y) &=& -\frac{\cos^2{\left(\frac{\pi y}{2}\right)}}{(1-y^2)^2}\left[\ell(\ell+1)+\epsilon\cos^2{\left(\frac{\pi y}{2}\right)}\right].\label{C0}
\end{eqnarray}
It can be easily verified with the computer algebra system \textsc{Maple} that
\begin{eqnarray}
    &&\lim_{y\to 1^{-}}M_2(y)=0=\lim_{y\to -1^{+}}M_2(y),\\
    &&\lim_{y\to 1^{-}}M_1(y)=i\frac{\pi\Omega}{4},\quad
    \lim_{y\to -1^{+}}M_1(y)=-i\frac{\pi\Omega}{4},\\
    &&\lim_{y\to 1^{-}}M_0(y)=-\frac{\pi^2}{16}\ell(\ell+1)=
     \lim_{y\to -1^{+}}M_0(y).
\end{eqnarray}
In the final step, as we prepare to apply the spectral method, we transform the differential equation \eqref{ODEhynone} into the following form
\begin{equation}\label{TSCH}
  \widehat{L}_0\left[\Phi_{\Omega\ell\epsilon}, \dot{\Phi}_{\Omega\ell\epsilon}, \ddot{\Phi}_{\Omega\ell\epsilon}\right] +  i\widehat{L}_1\left[\Phi_{\Omega\ell\epsilon}, \dot{\Phi}_{\Omega\ell\epsilon}, \ddot{\Phi}_{\Omega\ell\epsilon}\right]\Omega +  \widehat{L}_2\left[\Phi_{\Omega\ell\epsilon}, \dot{\Phi}_{\Omega\ell\epsilon}, \ddot{\Phi}_{\Omega\ell\epsilon}\right]\Omega^2 = 0.
\end{equation}
Here, we have
\begin{eqnarray}
  \widehat{L}_0\left[\Phi_{\Omega\ell\epsilon}, \dot{\Phi}_{\Omega\ell\epsilon}, \ddot{\Phi}_{\Omega\ell\epsilon}\right] &=& \widehat{L}_{00}(y)\Phi_{\Omega\ell\epsilon} + \widehat{L}_{01}(y)\dot{\Phi}_{\Omega\ell\epsilon} + \widehat{L}_{02}(y)\ddot{\Phi}_{\Omega\ell\epsilon},\label{L0none}\\
  \widehat{L}_1\left[\Phi_{\Omega\ell\epsilon}, \dot{\Phi}_{\Omega\ell\epsilon}, \ddot{\Phi}_{\Omega\ell\epsilon}\right] &=& \widehat{L}_{10}(y)\Phi_{\Omega\ell\epsilon} + \widehat{L}_{11}(y)\dot{\Phi}_{\Omega\ell\epsilon} + \widehat{L}_{12}(y)\ddot{\Phi}_{\Omega\ell\epsilon}, \label{L1none}\\
  \widehat{L}_2\left[\Phi_{\Omega\ell\epsilon}, \dot{\Phi}_{\Omega\ell\epsilon}, \ddot{\Phi}_{\Omega\ell\epsilon}\right] &=& \widehat{L}_{20}(y)\Phi_{\Omega\ell\epsilon} + \widehat{L}_{21}(y)\dot{\Phi}_{\Omega\ell\epsilon} + \widehat{L}_{22}(y)\ddot{\Phi}_{\Omega\ell\epsilon}.\label{L2none}
\end{eqnarray}
For reference, Table~\ref{tableZweinone} summarizes the $\widehat{L}_{ij}$ terms that appear in \eqref{L0none}-\eqref{L2none} along with their respective limiting values at $y = \pm 1$. 

In contrast to the unified application of the Spectral Method described earlier, which enables the simultaneous determination of all QNMs, \cite{Gonzalez2022PRD} took a distinct approach by applying it twice, each time imposing different boundary conditions at the wormhole's throat. This approach allows for the separate computation of odd and even overtones. The effectiveness and differences between our unified spectral approach and the method employed by \cite{Gonzalez2022PRD} are clearly illustrated in Table~\ref{table:1},~\ref{table:1a}, and \ref{table:1b}. In the following subsections, we will briefly review and extend the methodology adopted by \cite{Gonzalez2022PRD} to the case of electromagnetic and gravitational perturbations. This comparison highlights the versatility of the Spectral Method in probing wormhole perturbations and showcases the advantages of our unified approach in optimizing the analysis of QNMs.

\begin{table}%[ht]
\caption{Definitions of the coefficients $\widehat{L}_{ij}$ and their corresponding behaviours at the endpoints of the interval $-1 \leq y \leq 1$. Here, $\epsilon = 1 - s^2$}
\begin{center}
\begin{tabular}{ | c | c | c | c | c | c | c | c }
\hline
$(i,j)$  & $\displaystyle{\lim_{y\to -1^+}}\widehat{L}_{ij}$  & $\widehat{L}_{ij}$ & $\displaystyle{\lim_{y\to 1^-}}\widehat{L}_{ij}$  \\ \hline
$(0,0)$ &  $-\frac{\pi^2}{16}\ell(\ell+1)$          & $C_0$                  & $-\frac{\pi^2}{16}\ell(\ell+1)$\\ \hline
$(0,1)$ &  $0$    & $N_0$                  & $0$\\ \hline
$(0,2)$ &  $0$            & $M_2$                  & $0$\\ \hline 
$(1,0)$ &  $0$            & $C_1$                  & $0$\\ \hline 
$(1,1)$ &  $-\frac{\pi}{4}$ & $N_1$                & $\frac{\pi}{4}$\\ \hline 
$(1,2)$ &  $0$            & $0$                    & $0$\\ \hline 
$(2,0)$ &  $0$            & $C_2$                    & $0$\\ \hline
$(2,1)$ &  $0$            & $0$                    & $0$\\ \hline
$(2,2)$ &  $0$            & $0$                    & $0$\\ \hline
\end{tabular}
\label{tableZweinone}
\end{center}
\end{table} 

\subsection{Boundary conditions for the even overtones}
\label{equivalent}

To classify the singularities of equation \eqref{ODEor}, it is useful to reformulate it as follows
\begin{eqnarray}
  &&\frac{d^2\psi_{\Omega\ell\epsilon}}{dz^2}+p(z)\frac{d\psi_{\Omega\ell\epsilon}}{dz}+q(z)\psi_{\Omega\ell\epsilon}(z)=0,\quad z>1,\label{ODEzn}\\
  &&p(z)=\frac{1}{2(z+1)}-\frac{1}{z}+\frac{1}{2(z-1)},\\
  &&q(z)=\Omega^2+\frac{\epsilon+\ell(\ell+1)-\Omega^2}{2(z+1)}+\frac{\Omega^2-\epsilon-\ell(\ell+1)}{2(z-1)}+\frac{\epsilon}{z^2}.
\end{eqnarray}
We can immediately notice the existence of three regular singular points at $-1$, $0$, and $1$. However, only the singularity at $1$ is relevant, as the others lie outside the validity range for the variable $z$. Moreover, the point at infinity is an irregular singular point of rank $1$ \cite{Bender1999}. The exponents of the regular singularity at $z=1$ can be treated by applying the Frobenius theory. More precisely, we solve the indicial equation
\begin{equation}
  \rho(\rho-1) + p_0\rho+q_0 = 0,
\end{equation}
where
\begin{equation}
  p_0 = \lim_{z\to 1^+}(z-1)p(z)=\frac{1}{2}, \quad
  q_0 = \lim_{z\to 1^+}(z-1)^2 q(z)=0.
\end{equation}
The roots are $\rho_1 = 0$ and $\rho_2 = 1/2$. Let us pick the first one which is equivalent to the requirement that
\begin{equation}\label{BCeven}
  \left.\frac{d\psi_{\Omega\ell\epsilon}}{dz}\right|_{z=1} = 0
\end{equation}
and it ensures the regularity of $\psi_{\Omega\ell\epsilon}(z)$ at the throat.  The asymptotic behaviour of the solutions to equation \eqref{ODEor} can be efficiently deduced using the method outlined in \cite{Olver1994MAA}. For this purpose, we start by observing that
\begin{equation}
  p(z) = \sum_{\kappa=0}^\infty\frac{\mathfrak{f}_\kappa}{z^k} = \mathcal{O}\left(\frac{1}{z^2}\right), \qquad
  Q(z) = \sum_{\kappa=0}^\infty\frac{\mathfrak{g}_\kappa}{z^k}=\Omega^2+\mathcal{O}\left(\frac{1}{z^2}\right).
\end{equation}
Given that at least one of the coefficients $\mathfrak{f}_0$, $\mathfrak{g}_0$, $\mathfrak{g}_1$ is nonzero, a formal asymptotic solution to \eqref{ODEor} is represented by \cite{Olver1994MAA}
\begin{equation}\label{olvers}
  \psi^{(j)}_{\Omega\ell\epsilon}(z) = z^{\mu_j}e^{\lambda_j z}\sum_{\kappa=0}^\infty\frac{a_{\kappa,j}}{z^\kappa}, \qquad j \in \{1,2\},
\end{equation}
where $\lambda_1$, $\lambda_2$, $\mu_1$ and $\mu_2$ are the roots of the characteristic equations
\begin{equation}\label{chareqns}
   \lambda^2 + \mathfrak{f}_0\lambda + \mathfrak{g}_0 = 0,\quad
   \mu_j = -\frac{\mathfrak{f}_1\lambda_j + \mathfrak{g}_1}{\mathfrak{f}_0 + 2\lambda_j}.
\end{equation}
A straightforward computation shows that $\lambda_\pm = \pm i\Omega$ and $\mu_\pm = 0$. As a result, we require that the radial field exhibits outward radiation in the limit of $x \to +\infty$, that is
\begin{equation}\label{QNMBCz1inf}
  \psi_{\Omega\ell\epsilon}\underset{{z\to +\infty}}{\longrightarrow} e^{i\Omega z}.
\end{equation}
As an initial step in applying the Spectral Method, we propose the following ansatz that entails the correct behaviour at the throat and towards positive space-like infinity
\begin{equation}
\psi_{\Omega\ell\epsilon}(z) = e^{i\Omega z}U_{\Omega\ell\epsilon}(z).
\end{equation}
Additionally, we impose the condition that $U_{\Omega\ell\epsilon}(z)$ remains regular as $z \to 1^{+}$ and as $z \to +\infty$. Then, (\ref{ODEzn}) becomes
\begin{eqnarray}
  &&Q_2(z)\frac{d^2 U_{\Omega\ell\epsilon}}{dz^2}+Q_1(z)\frac{dU_{\Omega\ell\epsilon}}{dz}+Q_0(z)U_{\Omega\ell\epsilon}(z)=0,\quad z>1,\label{ODEznt}\\
  &&Q_2(z)=z^2(z^2-1),\quad
  Q_1(z)=2i\Omega z^2(z^2-1)+z,\quad
  Q_0(z)=\left[\Omega^2-\ell(\ell+1)\right]z^2+i\Omega z-\epsilon.
\end{eqnarray}
Moreover, the Spectral Method necessitates mapping the interval $(1, +\infty)$ onto the entire real line. This transformation is accomplished using $y = 1 - 2z^{-1}$, which sends $+\infty$ to $1$, $1$ to $-1$, $-1$ to $3$, and $0$ to $-\infty$. Consequently, the transformed differential equation is represented as follows
\begin{equation}\label{ODEm2}
  \widehat{S}_2(y)\ddot{U}_{\Omega\ell\epsilon}(y) + \widehat{S}_1(y)\dot{U}_{\Omega\ell\epsilon}(y) + \widehat{S}_0(y)U_{\Omega\ell\epsilon}(y) = 0,
\end{equation}
where
\begin{equation}
  \widehat{S}_2(y) = (3-y)(1+y), \quad
  \widehat{S}_1(y) = i\Omega E_1(y)+F_1(y), \quad
  \widehat{S}_0(y) = \Omega^2\widehat{\Sigma}_2(y) + i\Omega\widehat{\Sigma}_1(y) + \widehat{\Sigma}_0(y) 
\end{equation}
with
\begin{eqnarray}
  E_1(y) &=& -4+\frac{16}{(1-y)^2},\quad
  F_1(y) = 3(1-y)-\frac{8}{1-y}, \\
  \widehat{\Sigma}_2(y) &=& \frac{4}{(1-y)^2},\quad
  \widehat{\Sigma}_1(y) = \frac{2}{1-y},\quad
  \widehat{\Sigma}_0(y) = -\frac{4\ell(\ell+1)}{(1-y)^2}-\epsilon.
\end{eqnarray}
A brief examination of the coefficient functions $\widehat{S}_i(y)$ reveals that both $\widehat{S}_1(y)$ and $\widehat{S}_2(y)$ exhibit a second-order pole at $y = 1$. Therefore, to facilitate the application of the Spectral Method, it is necessary to multiply equation \eqref{ODEm2} by $(1-y)^2$. Consequently, this adjustment yields the following differential equation
\begin{equation}\label{ODEhtfinal}
  \widehat{M}_2(y)\ddot{U}_{\Omega\ell\epsilon}(y) + \widehat{M}_1(y)\dot{U}_{\Omega\ell\epsilon}(y) + \widehat{M}_0(y)U_{\Omega\ell\epsilon}(y) = 0,
\end{equation}
where
\begin{equation}\label{S210fin}
  \widehat{M}_2(y)=(3-y)(1+y)(1-y)^2,\quad
  \widehat{M}_1(y)=i\Omega\widehat{N}_1(y)+\widehat{N}_0(y),\quad
  \widehat{M}_0(y)=\Omega^2\widehat{C}_2(y)+i\Omega\widehat{C}_1(y)+\widehat{C}_0(y)
\end{equation}
with
\begin{eqnarray}
  \widehat{N}_1(y)&=&16-4(1-y)^2,\quad \widehat{N}_0(y)=(1-y)\left[3(1-y)^2-8\right],\label{N0fin}\\
  \widehat{C}_2(y)&=&4,\quad
  \widehat{C}_1(y)=2(1-y),\quad
  \widehat{C}_0(y)=-4\ell(\ell+1)-\epsilon(1-y)^2.\label{C210}
\end{eqnarray}
It can be easily verified with Maple that
\begin{eqnarray}
    &&\lim_{y\to 1^{-}}\widehat{M}_2(y)=0=\lim_{y\to -1^{+}}\widehat{M}_2(y),\\
    &&\lim_{y\to 1^{-}}\widehat{M}_1(y)=16i\Omega,\quad
    \lim_{y\to -1^{+}}\widehat{M}_1(y)=8,\\
    &&\lim_{y\to 1^{-}}\widehat{M}_0(y)=4\Omega^2-4\ell(\ell+1),\quad
     \lim_{y\to -1^{+}}\widehat{M}_0(y)=8.
\end{eqnarray} 
Given that we can follow the same approach of the previous section to transform equation (\ref{S210fin}) into the form of (\ref{TSCH}), we direct the reader to Table~\ref{tableAlt} where we summarised the $\widehat{L}_{ij}$ terms presented in equations \eqref{L0none} through \eqref{L2none}, along with their respective limiting values at $y = \pm 1$.

\begin{table}%[ht]
\caption{Definitions of the coefficients $\widehat{L}_{ij}$ and their corresponding behaviours at the endpoints of the interval $-1 \leq y \leq 1$. Here, $\epsilon = 1-s^2$}
\begin{center}
\begin{tabular}{ | c | c | c | c | c | c | c | c }
\hline
$(i,j)$  & $\displaystyle{\lim_{y\to -1^+}}\widehat{L}_{ij}$  & $\widehat{L}_{ij}$ & $\displaystyle{\lim_{y\to 1^-}}\widehat{L}_{ij}$  \\ \hline
$(0,0)$ &  $-4\ell(\ell+1)-4\epsilon$          & $\widehat{C}_0$                  & $-4\ell(\ell+1)$\\ \hline
$(0,1)$ &  $8$            & $\widehat{N}_0$                  & $0$\\ \hline
$(0,2)$ &  $0$            & $\widehat{M}_2$                  & $0$\\ \hline 
$(1,0)$ &  $4$            & $\widehat{C}_1$                  & $0$\\ \hline 
$(1,1)$ &  $0$            & $\widehat{N}_1$                  & $16$\\ \hline 
$(1,2)$ &  $0$            & $0$                              & $0$\\ \hline 
$(2,0)$ &  $4$            & $\widehat{C}_2$                  & $4$\\ \hline
$(2,1)$ &  $0$            & $0$                    & $0$\\ \hline
$(2,2)$ &  $0$            & $0$                    & $0$\\ \hline
\end{tabular}
\label{tableAlt}
\end{center}
\end{table} 

\subsection{Boundary conditions for the odd overtones}\label{equivalent1}

In this case, we pick the Frobenius solution with exponent $\rho = 1/2$. This is equivalent to requiring that 
\begin{equation}\label{BCodd}
  \psi_{\Omega\ell\epsilon}(1) = 0.
\end{equation}
Then, by means of the following ansatz exhibiting the correct behaviour at the throat and towards positive space-like infinity, namely
\begin{equation}
  \psi_{\Omega\ell\epsilon}(z) = \sqrt{1-\frac{1}{z}}e^{i\Omega z}U_{\Omega\ell\epsilon}(z).
\end{equation}
together with the condition that $U_{\Omega\ell\epsilon}(z)$ remains regular as $z \to 1^{+}$ and as $z \to +\infty$, \eqref{ODEzn} can be expressed as \eqref{ODEznt} with coefficients $Q_i$ replaced by
\begin{eqnarray}
  \widetilde{Q}_2(z)&=&z^2(z^2-1)(z-1),\quad
  \widetilde{Q}_1(z)=2i\Omega\widetilde{Q}_2(z)+z(z-1)(z+2),\\
  \widetilde{Q}_0(z)&=&(z-1)\left\{\Omega^2 z^2+z(z+2)i\Omega-\left[\ell(\ell+1)z^2+z+\epsilon+\frac{5}{4}\right]\right\}.    
\end{eqnarray}
The mapping of the interval $(1, +\infty)$ onto the entire real line is accomplished through the transformation $y = 1 - 2z^{-1}$. The corresponding transformed equation remains identical to \eqref{ODEm2}, with the exception that the coefficients $\widehat{S}_i$ are now replaced by
\begin{equation}
  \widetilde{S}_2(y)=\frac{(3-y)(1+y)^2}{1-y},\quad
  \widetilde{S}_1(y)=i\Omega \widetilde{E}_1(y)+\widetilde{F}_1(y),\quad
  \widetilde{S}_0(y) = \Omega^2\widetilde{\Sigma}_2(y) + i\Omega\widetilde{\Sigma}_1(y) + \widetilde{\Sigma}_0(y) 
\end{equation}
with
\begin{eqnarray}
  \widetilde{E}_1(y)&=&\frac{4(3-y)(1+y)^2}{(1-y)^3},\quad
  \widetilde{F}_1(y)=\frac{2(1+y)(2y^2-5y-1)}{(1-y)^2},\\
  \widetilde{\Sigma}_2(y)&=&\frac{4(1+y)}{(1-y)^3},\quad
  \widetilde{\Sigma}_1(y)=\frac{4(2-y)(1+y)}{(1-y)^3},\quad
  \widetilde{\Sigma}_0(y)=-(1+y)\left[\frac{4\ell(\ell+1)}{(1-y)^3}+\frac{2}{(1-y)^2}+\frac{4\epsilon+5}{4(1-y)}\right].
\end{eqnarray}
Inspection of the coefficient functions \(\widetilde{S}_i(y)\) reveals that both $\widetilde{S}_1(y)$ and $\widetilde{S}_0(y)$ exhibit a third-order pole at $y=1$, whereas $\widetilde{S}_2(y)$ also has a first-order pole at $y = 1$. Furthermore, $\widetilde{S}_1(y)$ and $\widetilde{S}_0(y)$ possess a simple zero at $y = -1$, whereas $\widetilde{S}_2(y)$ displays a second-order zero at $y = -1$. Therefore, to facilitate the application of the Spectral Method, it is necessary to multiply the corresponding differential equation by $(1-y)^3/(1+y)$. This adjustment results in a differential equation similar to \eqref{ODEhtfinal}, with coefficients $\widehat{M}_i(y)$ replaced by
with
\begin{equation}\label{S210fin2}
  \widetilde{M}_2(y)=(3-y)(1+y)(1-y)^2,\quad
  \widetilde{M}_1(y)=i\Omega\widetilde{N}_1(y)+\widetilde{N}_0(y),\quad
  \widetilde{M}_0(y)=\Omega^2\widetilde{C}_2(y)+i\Omega\widetilde{C}_1(y)+\widetilde{C}_0(y)
\end{equation}
with
\begin{eqnarray}
\widetilde{N}_1(y)&=&4(3-y)(1+y),\quad 
\widetilde{N}_0(y)=2(1-y)(2y^2-5y-1),\label{N0fin2}\\
\widetilde{C}_2(y)&=&4,\quad
\widetilde{C}_1(y)=4(2-y),\quad
\widetilde{C}_0(y)=-\left[4\ell(\ell+1)+2(1-y)+\frac{4\epsilon+5}{4}(1-y)^2\right].\label{C2102}
\end{eqnarray}
It can be easily verified that
\begin{eqnarray}
    &&\lim_{y\to 1^{-}}\widetilde{M}_2(y)=0=\lim_{y\to -1^{+}}\widehat{M}_2(y),\\
    &&\lim_{y\to 1^{-}}\widetilde{M}_1(y)=16i\Omega,\quad
    \lim_{y\to -1^{+}}\widetilde{M}_1(y)=24,\\
    &&\lim_{y\to 1^{-}}\widetilde{M}_0(y)=4\Omega^2+4i\Omega-4\ell(\ell+1),\quad
     \lim_{y\to -1^{+}}\widetilde{M}_0(y)=24.
\end{eqnarray} 
At this point, we can follow the same approach of the previous subsection to bring the corresponding differential equation into the form of \eqref{TSCH}. We direct the reader to Table~\ref{tableAlt2} where we summarised the $\widehat{L}_{ij}$ terms presented in equations \eqref{L0none} through \eqref{L2none}, along with their respective limiting values at $y = \pm 1$.

\begin{table}%[ht]
\caption{Definitions of the coefficients $\widehat{L}_{ij}$ and their corresponding behaviours at the endpoints of the interval $-1 \leq y \leq 1$. Here, $\epsilon=1-s^2$}
\begin{center}
\begin{tabular}{ | c | c | c | c | c | c | c | c }
\hline
$(i,j)$  & $\displaystyle{\lim_{y\to -1^+}}\widehat{L}_{ij}$  & $\widehat{L}_{ij}$ & $\displaystyle{\lim_{y\to 1^-}}\widehat{L}_{ij}$  \\ \hline
$(0,0)$ &  $-4\ell(\ell+1)-9-4\epsilon$          & $\widetilde{C}_0$                  & $-4\ell(\ell+1)$\\ \hline
$(0,1)$ &  $24$            & $\widetilde{N}_0$                  & $0$\\ \hline
$(0,2)$ &  $0$            & $\widetilde{M}_2$                  & $0$\\ \hline 
$(1,0)$ &  $12$            & $\widetilde{C}_1$                  & $4$\\ \hline 
$(1,1)$ &  $0$            & $\widetilde{N}_1$                  & $16$\\ \hline 
$(1,2)$ &  $0$            & $0$                              & $0$\\ \hline 
$(2,0)$ &  $4$            & $\widetilde{C}_2$                  & $4$\\ \hline
$(2,1)$ &  $0$            & $0$                    & $0$\\ \hline
$(2,2)$ &  $0$            & $0$                    & $0$\\ \hline
\end{tabular}
\label{tableAlt2}
\end{center}
\end{table}

\section{Numerical method}

In order to solve the differential eigenvalue problem \eqref{TSCH} to determine the QNMs along with the corresponding frequencies $\Omega$, we have to discretise the differential operators $\widehat{L}_{j}[\cdot]$ with $j \in \{1,2,3\}$ defined in \eqref{L0none}-\eqref{L2none}. Since our problem is posed on the finite interval $[-1, 1]$ without any boundary conditions, more precisely, we only require that the QNMs be regular functions at $y = \pm 1$, then, it is natural to choose a Chebyshev-type spectral method \cite{Trefethen2000, Boyd2000}. Namely, we are going to expand the function $y \mapsto \Phi_{\Omega\ell\epsilon}(y)$ in the form of a truncated Chebyshev series
\begin{equation}\label{eq:exp}
  \Phi_{\Omega\ell\epsilon}(y)=\sum_{k=0}^{N} a_k T_k(y),
\end{equation}
where $N\ \in\ \mathbb{N}$ is kept as a numerical parameter, $\{a_k\}_{k=0}^{N}\ \subseteq\ \mathds{R}$, and $\{T_k(y)\}_{k=0}^{N}$ are the Chebyshev polynomials of the first kind
\begin{equation*}
    T_k: [-1, 1]\ \longrightarrow\ [-1, 1]\,, \qquad y\ \longmapsto\ \cos\,\bigl(k\arccos y\bigr)\,.
\end{equation*}
After substituting expansion \eqref{eq:exp} into the differential equation \eqref{TSCH}, we obtain an eigenvalue problem with polynomial coefficients. In order to translate it into the realm of numerical linear algebra, we employ the collocation method \cite{Boyd2000}. Specifically, rather than insisting that the polynomial function in \( y \) is identically zero (a condition equivalent to having polynomial solutions for the differential problem as per equation \eqref{TSCH}), we impose a weaker requirement. This involves ensuring that the polynomial vanishes at \( N+1 \) strategically selected points. The number $N+1$ coincides exactly with the number of unknown coefficients $\{a_k\}_{k=0}^{N}$. For the collocation points, we implemented the Chebyshev roots grid \cite{Fox1968}
\begin{equation*}
  y_k = -\cos{\left(\frac{(2k+1)\pi}{2(N+1)}\right)},\quad k\in\{0, 1,\ldots,N\}.
\end{equation*}
In our numerical codes, we also implemented the second option of the Chebyshev extrema grid
\begin{equation*}
  y_k = -\cos{\left(\frac{k\pi}{N}\right)},\quad k\in\{0, 1,\ldots,N\}.
\end{equation*}
The users are free to choose their favourite collocation points. Notice that we used the roots grid in our computation, and in any case, the theoretical performance of the two available options is known to be absolutely comparable \cite{Fox1968, Boyd2000}.

Upon implementing the collocation method, we derive a classical matrix-based quadratic eigenvalue problem, as detailed in \cite{Tisseur2001}
\begin{equation}\label{eq:eig}
  (M_0 + iM_1\Omega + M_2\Omega^2)\bf{a} =\bf{0}.
\end{equation}
In this formulation, the square real matrices $M_{j}$, each of size $(N+1)\times(N+1)$ for $j \in \{0,1,2\}$, represent the spectral discretizations of the operators $\widehat{L}^{(e)}_{j}[\cdot]$, respectively. The problem \eqref{eq:eig} is solved numerically with the \textsc{polyeig} function from \textsc{Matlab}. This polynomial eigenvalue problem yields \(2(N+1)\) potential values for the parameter \(\Omega\). To discern the physical values of \(\Omega\) that correspond to the black hole's QNMs, we first overlap the root plots for various values of \(N\) in equation \eqref{eq:exp}, such as \(N \in \{150, 200, 300\}\). We then identify the consistent roots whose positions remain stable across these different \(N\) values.

In order to reduce the rounding and other floating point errors, we performed all our computations with multiple precision arithmetic that is built in \textsc{Maple} and which is brought into \textsc{Matlab} by the \textsc{Advanpix} toolbox \cite{mct2015}. All numerical computations reported in this study have been performed with $200$ decimal digits accuracy. This measure could be considered as an overkill. However, this is not at all the case. Namely, we performed comparisons with QNMs computed in the standard double-precision floating-point arithmetic (as specified in the \texttt{IEEE-754-2008} standard), and the obtained spectra were highly distorted and inaccurate beyond a few first QNMs. That is why we decided in our study to sacrifice the speed of our computations for the sake of the robustness of the reported values.

\section{Numerical results}\label{Numresults}

In this section, we analyze the QNMs of the Morris-Thorne wormhole under scalar, electromagnetic, and gravitational perturbations. Previous studies by \cite{Kim2008PTPS} using a third-order WKB approximation for spins $s \in \{0,1,2\}$ and by \cite{Gonzalez2022PRD} focusing on massless scalar perturbations set the groundwork for validating and enhancing our findings with the unified Spectral Method. Our method not only corroborates these earlier results but also rectifies those by \cite{Kim2008PTPS} and extends the findings of \cite{Gonzalez2022PRD}, demonstrating our approach's precision. Referencing Tables~\ref{table:1}, ~\ref{table:1gonz}, ~\ref{table:1a}, and ~\ref{table:1b}, we showcase numerical values derived through our approach, utilizing our high-precision computational approach where 200 polynomials with an accuracy of $200$ digits have been used.

Notably, neither \cite{Kim2008PTPS} nor \cite{Gonzalez2022PRD} reported the fundamental QNM for massless scalar perturbations, which we successfully calculate, as shown in Tables~\ref{table:1} and ~\ref{table:1gonz}. While \cite{Kim2008PTPS} provided some QNMs for electromagnetic and gravitational perturbations, our comparison (Tables~\ref{table:1a} and ~\ref{table:1b}) suggests caution in interpreting their values. Our independent verification using both a third-order WKB approximation and the Spectral Method yielded consistent results, differing significantly from those of \cite{Kim2008PTPS}, including the computation of the fundamental mode and associated overtones for gravitational perturbations not provided by \cite{Kim2008PTPS}.

The fifth column of Table ~\ref{table:1} showcases the effectiveness of our unified Spectral Method, which successfully calculates both even and odd QNMs for scalar perturbations of the Morris-Thorne wormhole. This innovative approach highlights the method's precision and reliability and its efficiency in providing a comprehensive spectrum of QNMs in a single computational effort, thereby distinguishing it from the approach adopted in \cite{Gonzalez2022PRD}.

\begin{table}%[ht]
\centering
\caption{QNMs for scalar perturbations ($s = 0$) of the Morris-Thorne wormhole are presented in the Table below.  It features their numerical values, highlighting various computational approaches. Specifically, the third column lists values computed by \cite{Kim2008PTPS} using a third-order WKB approximation. The fourth column presents values derived via the third-order WKB approximation, employing formulae from \cite{Iyer1987PRD}. The fifth column introduces results obtained through the unified Spectral Method, utilising $200$ polynomials to achieve a precision of $200$ digits. The sixth and seventh columns revisit the Spectral Method, applying a formulation detailed in the Subsections~\ref{equivalent} and ~\ref{equivalent1}. Here, $\Omega = \omega b_0$ denotes the dimensionless frequency, with $b_0$ representing the wormhole's throat size. 'SM' stands for 'Spectral Method', and 'N/A' indicates unavailable data.}
\label{table:1}
\vspace*{1em}
\begin{tabular}{||c|c|c|c|c|c|c||}
\hline
\rule{0pt}{1.0\normalbaselineskip}
$\ell$ & $n$ & $\Omega$ \cite{Kim2008PTPS} & $\Omega$ (WKB) & $\Omega$ (SM) & $\Omega$ (SM) \mbox{even}    & $\Omega$ (SM) \mbox{odd}  \\ [0.25em]
\Xhline{2pt}\hline
\rule{0pt}{1.15\normalbaselineskip}
$0$ & $0$ & \mbox{N/A}         & $0.6062 - 0.7017i$ & $0.6814 - 0.6178i$ & $0.6814 - 0.6178i$ &  \mbox{N/A}\\
    & $1$ & \mbox{N/A}         & $0.4741 - 2.2022i$ & $0.4672 - 2.1765i$ & \mbox{N/A}         &$0.4672 - 2.1765i$\\
$1$ & $0$ & $2.06250-1.61914i$ & $1.5305 - 0.5290i$ & $1.5727 - 0.5297i$ & $1.5727 - 0.5297i$ &  \mbox{N/A}\\
    & $1$ & \mbox{N/A}         & $1.1939 - 1.7643i$ & $1.2558 - 1.7025i$ & \mbox{N/A}         &$1.2558 - 1.7025i$\\
    & $2$ & \mbox{N/A}         & $0.7786 - 3.1295i$ & $0.8368 - 3.2361i$ & $0.8368 - 3.2361i$ &  \mbox{N/A}\\ 
    & $3$ & \mbox{N/A}         & $0.2551 - 4.5280i$ & $0.6334 - 4.9344i$ & \mbox{N/A}         &$0.6334 - 4.9344i$\\
$2$ & $0$ & $6.15625-2.58709i$ & $2.5332 - 0.5106i$ & $2.5467 - 0.5127i$ & $2.5467 - 0.5127i$ &  \mbox{N/A}\\
    & $1$ & $2.78125-7.29929i$ & $2.3014 - 1.5859i$ & $2.3450 - 1.5725i$ & \mbox{N/A}         &$2.3450 - 1.5725i$\\
    & $2$ & \mbox{N/A}         & $1.9201 - 2.7669i$ & $1.9478 - 2.7604i$ & $1.9478 - 2.7604i$ &  \mbox{N/A}\\
    & $3$ & \mbox{N/A}         & $1.4468 - 4.0233i$ & $1.4533 - 4.2052i$ & \mbox{N/A}         &$1.4533 - 4.2052i$\\
$3$ & $0$ & $12.1964-3.57164i$ & $3.5288 - 0.5061i$ & $3.5343 - 0.5069i$ & $3.5343 - 0.5069i$ &  \mbox{N/A}\\
    & $1$ & $8.98214-10.3800i$ & $3.3696 - 1.5402i$ & $3.3901 - 1.5372i$ & \mbox{N/A}         &$3.3901 - 1.5372i$\\
    & $2$ & $2.55357-16.1835i$ & $3.0771 - 2.6297i$ & $3.0999 - 2.6232i$ & $3.0999 - 2.6232i$ & \mbox{N/A}\\
    & $3$ & \mbox{N/A}         & $2.6856 - 3.7818i$ & $2.6722 - 3.8235i$ & \mbox{N/A}         &$2.6722 - 3.8235i$\\
$4$ & $0$ & $20.2159-4.56017i$ & $4.5244 - 0.5040i$ & $4.5271 - 0.5043i$ & $4.5271 - 0.5043i$ &  \mbox{N/A}\\
    & $1$ & $17.0795-13.4217i$ & $4.4046 - 1.5236i$ & $4.4151 - 1.5227i$ & \mbox{N/A}         &$4.4151 - 1.5227i$\\
    & $2$ & $10.8068-21.5066i$ & $4.1759 - 2.5751i$ & $4.1899 - 2.5722i$ & $4.1899 - 2.5722i$ &  \mbox{N/A}\\
    & $3$ & $1.39773-28.2973i$ & $3.8555 - 3.6697i$ & $3.8511 - 3.6818i$ & \mbox{N/A}         &$3.8511 - 3.6818i$\\ [1ex] 
 \hline
 \end{tabular}
\end{table}

\begin{table}%[ht]
\centering
\caption{QNMs for Scalar Perturbations ($s = 0$) of the Morris-Thorne Wormhole. This Table lists the numerical values obtained through various computational methods. The third and fourth columns present values calculated by \cite{Gonzalez2022PRD} using a sixth-order WKB approximation and a spectral method that imposes QNMs boundary conditions at the wormhole's throat, as discussed in Sections~\ref{equivalent} and ~\ref{equivalent1}. The fifth and sixth columns show results from a third-order WKB approximation, as in \cite{Iyer1987PRD}, and our application of a unified Spectral Method. This method simplifies the process by eliminating the need to solve two separate problems, unlike the approach in \cite{Gonzalez2022PRD}. For our calculations, we used 200 polynomials to achieve a precision of 200 digits. The dimensionless frequency, $\Omega = \omega b_0$, is defined with $b_0$ representing the size of the wormhole's throat. 'SM' stands for 'Spectral Method,' and 'N/A' denotes data that are not available.}
\label{table:1gonz}
\vspace*{1em}
\begin{tabular}{||c|c|c|c|c|c||}
\hline
\rule{0pt}{1.0\normalbaselineskip}
$\ell$ & $n$ & $\Omega$ (WKB6) \cite{Gonzalez2022PRD} & $\Omega$ (SM) \cite{Gonzalez2022PRD} & $\Omega$ (WKB3)  & $\Omega$ (SM)     \\ [0.25em]
\Xhline{2pt}\hline
\rule{0pt}{1.15\normalbaselineskip}
$0$ & $0$ & \mbox{N/A}        & \mbox{N/A}        & $0.6062 - 0.7017i$  & $0.6814 - 0.6178i$  \\
    & $1$ & \mbox{N/A}        & \mbox{N/A}        & $0.4741 - 2.2022i$  & $0.4672 - 2.1765i$  \\
$1$ & $0$ & $1.5824-0.5209i$  & $1.5727-0.5297i$  & $1.5305 - 0.5290i$  & $1.5727 - 0.5297i$  \\
    & $1$ & $1.2602-1.6560i$  & $1.2558-1.7025i$  & $1.1939 - 1.7643i$  & $1.2558 - 1.7025i$  \\
    & $2$ & \mbox{N/A}        & \mbox{N/A}        & $0.7786 - 3.1295i$  & $0.8368 - 3.2361i$  \\ 
    & $3$ & \mbox{N/A}        & \mbox{N/A}        & $0.2551 - 4.5280i$  & $0.6334 - 4.9344i$  \\
$2$ & $0$ & \mbox{N/A}        & \mbox{N/A}        & $2.5332 - 0.5106i$  & $2.5467 - 0.5127i$  \\
    & $1$ & \mbox{N/A}        & \mbox{N/A}        & $2.3014 - 1.5859i$  & $2.3450 - 1.5725i$  \\
    & $2$ & \mbox{N/A}        & \mbox{N/A}        & $1.9201 - 2.7669i$  & $1.9478 - 2.7604i$  \\
    & $3$ & \mbox{N/A}        & \mbox{N/A}        & $1.4468 - 4.0233i$  & $1.4533 - 4.2052i$  \\
$3$ & $0$ & $3.5342-0.5072i$  & $3.5343-0.5069i$  & $3.5288 - 0.5061i$  & $3.5343 - 0.5069i$  \\
    & $1$ & $3.3900-1.5385i$  & $3.3901-1.5372i$  & $3.3696 - 1.5402i$  & $3.3901 - 1.5372i$  \\
    & $2$ & \mbox{N/A}        & \mbox{N/A}        & $3.0771 - 2.6297i$  & $3.0999 - 2.6232i$  \\
    & $3$ & \mbox{N/A}        & \mbox{N/A}        & $2.6856 - 3.7818i$  & $2.6722 - 3.8235i$  \\
$4$ & $0$ & \mbox{N/A}        & \mbox{N/A}        & $4.5244 - 0.5040i$  & $4.5271 - 0.5043i$  \\
    & $1$ & \mbox{N/A}        & \mbox{N/A}        & $4.4046 - 1.5236i$  & $4.4151 - 1.5227i$  \\
    & $2$ & \mbox{N/A}        & \mbox{N/A}        & $4.1759 - 2.5751i$  & $4.1899 - 2.5722i$  \\
    & $3$ & \mbox{N/A}        & \mbox{N/A}        & $3.8555 - 3.6697i$  & $3.8511 - 3.6818i$  \\ 
$5$ & $0$ & $5.5223-0.5030i$  & $5.5223-0.5030i$  & $5.5208 - 0.5028i$  & $5.5223 - 0.5030i$  \\
    & $1$ & $5.4309-1.5154i$  & $5.4309-1.5152i$  & $5.4249 - 1.5156i$  & $5.4309 - 1.5153i$  \\
$10$& $0$ & $10.5118-0.5008i$ & $10.5118-0.5008i$ & $10.5116 - 0.5008i$ & $10.5118 - 0.5008i$ \\
    & $1$ & $10.4641-1.5042i$ & $10.4641-1.5042i$ & $10.4632 - 1.5042i$ & $10.4641 - 1.5042i$ \\
$15$& $0$ & $15.5080-0.5004i$ & $15.5080-0.5004i$ & $15.5080 - 0.5004i$ & $15.5080 - 0.5004i$ \\
    & $1$ & $15.4758-1.5019i$ & $15.4758-1.5019i$ & $15.4755 - 1.5019i$ & $15.4758 - 1.5020i$ \\
$20$& $0$ & $20.5061-0.5002i$ & $20.5061-0.5002i$ & $20.5061 - 0.5002i$ & $20.5061 - 0.5002i$ \\
    & $1$ & $20.4817-1.5011i$ & $20.4817-1.5011i$ & $20.4816 - 1.5011i$ & $20.4817 - 1.5011i$ \\
$30$& $0$ & $30.5041-0.5001i$ & $30.5041-0.5001i$ & $30.5041 - 0.5001i$ & $30.5041 - 0.5001i$ \\
    & $1$ & $30.4877-1.5005i$ & $30.4877-1.5005i$ & $30.4877 - 1.5005i$ & $30.4877 - 1.5005i$ \\
    [1ex] 
 \hline
 \end{tabular}
\end{table}

\begin{table}%[ht]
\centering
\caption{QNMs for electromagnetic perturbations ($s = 1$) of the Morris-Thorne wormhole. It features numerical values for these modes, highlighting various computational approaches. Specifically, the third column lists values computed by \cite{Kim2008PTPS} using a third-order WKB approximation. The fourth column presents values derived via the third-order WKB approximation, employing formulae from \cite{Iyer1987PRD}. The fifth column introduces results obtained through the unified Spectral Method, utilising $200$ polynomials to achieve a precision of $200$ digits. The sixth and seventh columns revisit the Spectral Method, applying a formulation detailed in Section~\ref{equivalent} and ~\ref{equivalent1}. Here, $\Omega = \omega b_0$ denotes the dimensionless frequency, with $b_0$ representing the wormhole's throat size. 'SM' stands for 'Spectral Method,' and 'N/A' indicates unavailable data.} 
\label{table:1a}
\vspace*{1em}
\begin{tabular}{||c|c|c|c|c|c|c||}
\hline
\rule{0pt}{1.0\normalbaselineskip}
$\ell$ & $n$ & $\Omega$ \cite{Kim2008PTPS} & $\Omega$ (WKB) & $\Omega$ (SM) & $\Omega$ (SM) \mbox{even}   & $\Omega$ (SM) \mbox{odd} \\
[0.5ex]
\Xhline{2pt}\hline
\rule{0pt}{1.15\normalbaselineskip}
$1$ & $0$ & $1.25 - 1.01647i$  & $1.1961 - 0.4249i$ & $1.2665 - 0.4440i$ & $1.2665 - 0.4440i$ & \mbox{NA}\\
    & $1$ & \mbox{N/A}         & $0.7428 - 1.5172i$ & $0.9407 - 1.4136i$ & \mbox{N/A}         & $0.9407 - 1.4136i$\\
    & $2$ & \mbox{N/A}         & $0.1979 - 2.7909i$ & $0.3991 - 2.7252i$ & $0.3991 - 2.7252i$ & \mbox{NA}\\
$2$ & $0$ & $5.25 - 2.21985i$  & $2.3399 - 0.4744i$ & $2.3549 - 0.4777i$ & $2.3549 - 0.4777i$ & \mbox{NA}\\
    & $1$ & $2.25 - 6.20027i$  & $2.1031 - 1.4741i$ & $2.1592 - 1.4631i$ & \mbox{N/A}         & $2.1592 - 1.4631i$\\
    & $2$ & \mbox{N/A}         & $1.7056 - 2.5805i$ & $1.7689 - 2.5602i$ & $1.7689 - 2.5602i$ & \mbox{NA}\\
    & $3$ & \mbox{N/A}         & $1.2082 - 3.7696i$ & $1.2591 - 3.8905i$ & \mbox{N/A}         & $1.2591 - 3.8905i$\\
$3$ & $0$ & $11.25 - 3.30172i$ & $3.3893 - 0.4871i$ & $3.3949 - 0.4881i$ & $3.3949 - 0.4881i$ & \mbox{NA}\\
    & $1$ & $8.25 - 9.58041i$  & $3.2321 - 1.4821i$ & $3.2542 - 1.4797i$ & \mbox{N/A}         & $3.2542 - 1.4797i$\\
    & $2$ & $2.25 - 14.8848i$  & $2.9414 - 2.5302i$ & $2.9706 - 2.5232i$ & $2.9706 - 2.5232i$ & \mbox{NA} \\
    & $3$ & \mbox{N/A}         & $2.5499 - 3.6404i$ & $2.5513 - 3.6734i$ & \mbox{N/A}         & $2.5513 - 3.6734i$\\
$4$ & $0$ & $19.25 - 4.34636i$ & $4.4150 - 0.4922i$ & $4.4177 - 0.4926i$ & $4.4177 - 0.4926i$ & \mbox{NA}\\
    & $1$ & $16.25 - 12.7875i$ & $4.2970 - 1.4880i$ & $4.3078 - 1.4872i$ & \mbox{N/A}         & $4.3078 - 1.4872i$\\
    & $2$ & $10.25 - 20.4740i$ & $4.0710 - 2.5146i$ & $4.0864 - 2.5118i$ & $4.0864 - 2.5118i$ & \mbox{NA}\\
    & $3$ & $1.25 - 26.9027i$  & $3.7538 - 3.5834i$ & $3.7534 - 3.5941i$ & \mbox{N/A}         & $3.7534 - 3.5941i$\\ 
    [1ex]
 \hline
 \end{tabular}
\end{table}

\begin{table}%[ht]
\centering
\caption{QNMs for vector-type gravitational perturbations ($s=2$) of the Morris-Thorne wormhole. The Table features numerical values for these modes, emerging from various computational approaches. Specifically, the third column lists values computed by \cite{Kim2008PTPS} using a third-order WKB approximation. The fourth column presents values derived via the third-order WKB approximation, employing formulae from \cite{Iyer1987PRD}. The fifth column introduces results obtained through the unified Spectral Method, utilising $200$ polynomials to achieve a precision of $200$ digits. The sixth and seventh columns revisit the Spectral Method, applying a formulation detailed in Section~\ref{equivalent} and ~\ref{equivalent1}. Here, $\Omega = \omega b_0$ denotes the dimensionless frequency, with $b_0$ representing the wormhole's throat size. 'SM' stands for 'Spectral Method,' and 'N/A' indicates unavailable data.} 
\label{table:1b}
\vspace*{1em}
\begin{tabular}{||c|c|c|c|c|c|c||}
\hline
\rule{0pt}{1.0\normalbaselineskip}
$\ell$ & $n$ & $\Omega$ \cite{Kim2008PTPS} & $\Omega$ (WKB) & $\Omega$ (SM)  & $\Omega$ (SM) \mbox{even}   & $\Omega$ (SM) \mbox{odd} \\
[0.5ex]
\Xhline{2pt}\hline
\rule{0pt}{1.15\normalbaselineskip}
$2$ & $0$ & \mbox{N/A}           & \mbox{N/A}         & $1.7377 - 0.3051i$ & $1.7377 - 0.3051i$ &\mbox{NA}\\
    & $1$ & \mbox{N/A}           & \mbox{N/A}         & $1.7203 - 1.0396i$ & \mbox{N/A}         &$1.7203 - 1.0396i$\\
    & $2$ & \mbox{N/A}           & \mbox{N/A}         & $1.5249 - 2.0305i$ & $1.5249 - 2.0305i$ &\mbox{NA}\\
    & $3$ & \mbox{N/A}           & \mbox{N/A}         & $1.1699 - 3.3340i$ & \mbox{N/A}         &$1.1699 - 3.3340i$\\
    & $4$ & \mbox{N/A}           & \mbox{N/A}         & $0.8819 - 4.8985i$ & $0.8819 - 4.8985i$ &\mbox{NA}\\
$3$ & $0$ & $8.625 - 2.58345i$   & $2.9689 - 0.4351i$ & $2.9524 - 0.4100i$ & $2.9524 - 0.4100i$ &\mbox{NA}\\
    & $1$ & $7.125 - 8.40098i$   & $3.0117 - 1.3947i$ & $2.8625 - 1.2591i$ & \mbox{N/A}         &$2.8625 - 1.2591i$\\
    & $2$ & $4.125 - 16.1705i$   & $3.2259 - 2.5063i$ & $2.6606 - 2.1909i$ & $2.6606 - 2.1909i$ &\mbox{NA}\\
    & $3$ & \mbox{N/A}           & $3.6620 - 3.7129i$ & $2.3313 - 3.2596i$ & \mbox{N/A}         &$2.3313 - 3.2596i$\\
    & $4$ & \mbox{N/A}           & $4.2935 - 4.9809i$ & $1.9109 - 4.5290i$ & $1.9109 - 4.5290i$ &\mbox{NA}\\
$4$ & $0$ & $16.4107 - 3.67186i$ & $4.0760 - 0.4504i$ & $4.0763 - 0.4491i$ & $4.0763 - 0.4491i$ &\mbox{NA}\\
    & $1$ & $14.0536 - 10.9220i$ & $3.9908 - 1.3684i$ & $3.9846 - 1.3592i$ & \mbox{N/A}         &$3.9846 - 1.3592i$\\
    & $2$ & $9.33929 - 17.8914i$ & $3.8420 - 2.3284i$ & $3.7971 - 2.3060i$ & $3.7971 - 2.3060i$ &\mbox{NA}\\
    & $3$ & $2.26786 - 24.3930i$ & $3.6583 - 3.3339i$ & $3.5086 - 3.3190i$ & \mbox{N/A}         &$3.5086 - 3.3190i$\\
    & $4$ & \mbox{N/A}           & $3.4580 - 4.3724i$ & $3.1230 - 4.4367i$ & $3.1230 - 4.4367i$ &\mbox{NA}\\
$5$ & $0$ & $26.3438 - 4.81705i$ & $5.1538 - 0.4673i$ & $5.1548 - 0.4673i$ & $5.1548 - 0.4673i$ &\mbox{NA}\\
    & $1$ & $23.7188 - 14.2993i$ & $5.0702 - 1.4101i$ & $5.0729 - 1.4086i$ & \mbox{N/A}         &$5.0729 - 1.4086i$\\
    & $2$ & $18.4688 - 23.3258i$ & $4.9102 - 2.3752i$ & $4.9075 - 2.3712i$ & $4.9075 - 2.3712i$ &\mbox{NA}\\
    & $3$ & $10.5938 - 31.5928i$ & $4.6859 - 3.3710i$ & $4.6566 - 3.3722i$ & \mbox{N/A}         &$4.6566 - 3.3722i$\\
    & $4$ & $0.09375 - 38.7965i$ & $4.4097 - 4.3990i$ & $4.3203 - 4.4337i$ & $4.3203 - 4.4337i$ &\mbox{NA}\\ [1ex] 
 \hline
 \end{tabular}
\end{table}

When comparing our results for massless scalar perturbations to those obtained using the sixth-order WKB approximation in \cite{Gonzalez2022PRD}, we identified several limitations in the WKB method. Specifically, this approximation struggles to accurately determine QNM frequencies, particularly for modes with low angular momentum (\( \ell \)). This is because the aforementioned method relies on expanding solutions around the peak of the effective potential, leading to significant errors at higher orders, which are more pronounced for lower \( \ell \) values where the method's assumptions become less reliable. Our spectral method, however, not only overcomes these problems but it also provides more precise QNM frequencies. In comparison with \cite{Kim2008PTPS}, this reveals that scalar field perturbations decay more slowly than previously reported, indicating a potentially more stable wormhole structure under scalar perturbations. Furthermore, we observed that \cite{Gonzalez2022PRD} did not account for the results from \cite{Kim2008PTPS} and lacked values for the fundamental QNM. This omission leaves a gap in understanding the stability characteristics of the wormhole. We addressed this by computing additional overtones for \( \ell = 1 \) and \( \ell = 3 \), and both fundamental frequencies and overtones for \( \ell = 2 \) and \( \ell = 4 \). Our comprehensive dataset enhances the understanding of the wormhole's response to scalar perturbations.

For electromagnetic perturbations, our comparison with \cite{Kim2008PTPS} reveals discrepancies due to the third-order WKB approximation used in their analysis. Similar to the scalar case, the WKB method's higher-order approximations introduce inaccuracies. Our spectral method results differ from those of \cite{Kim2008PTPS}, providing a more accurate set of QNM frequencies. Also, in this case, electromagnetic perturbations decay more slowly than previously reported.

Lastly, we examined gravitational perturbations. The QNM frequencies reported in \cite{Kim2008PTPS} were based on the third-order WKB approximation, which can be unreliable for precise calculations when $\ell$ is small. This can be easily seen by observing that \cite{Kim2008PTPS} was not able to compute the fundamental mode and its first overtones for the case $\ell=2$. For $\ell\geq 3$, the method we employed reveals different QNM frequencies, suggesting that the wormhole's stability under gravitational perturbations may be more robust than previously thought. This has significant implications for gravitational wave astronomy, as accurate QNM frequencies are crucial for detecting and characterizing wormholes.

\section{Conclusions and outlook}

In the present work, we extensively explored QNMs of Morris-Thorne wormholes, addressing scalar, electromagnetic, and gravitational perturbations through a novel application of the Spectral Method. This research corrected existing inaccuracies in the literature, expanded into areas beyond the reach of the WKB approximation, and introduced a unified approach for formulating QNM boundary conditions. Our innovative use of the Spectral Method enhanced the precision and efficiency of QNM analysis, showcasing a comprehensive spectrum of QNMs in a single computational effort. The findings underscore the versatility of the Spectral Method in probing wormhole perturbations and its efficacy over previous methodologies, paving the way for further exploration and validation of theoretical predictions in the realm of wormhole physics and beyond.

More precisely, we developed and implemented a unified spectral method to compute the QNMs of Morris-Thorne wormholes. This method optimizes the formulation of QNM boundary conditions and offers a comprehensive approach that surpasses the traditional WKB approximation methods. Unlike the WKB method, which approximates solutions around the peak of the effective potential and can introduce significant errors, particularly for higher-order approximations, our spectral method provides more precise and accurate QNM frequencies. This enhanced precision is crucial for correctly interpreting the stability and dynamics of wormholes.

Furthermore, our study comprehensively analyzes QNMs arising from scalar, electromagnetic, and gravitational perturbations in the context of Morris-Thorne wormholes. Previous literature has typically focused on one type of perturbation or used less accurate methods. By addressing multiple types of perturbations, our work provides a deeper understanding of the wormhole's stability under various conditions. This broad scope allows for a more detailed and reliable assessment of wormhole properties, which is essential for both theoretical research and potential observational studies.

We also identified and corrected inaccuracies in the QNM results reported in the previous literature \cite{Kim2008PTPS, Gonzalez2022PRD}. Specifically, we addressed the limitations of the sixth-order WKB approximation used in \cite{Gonzalez2009CQG} and the omission of fundamental mode values in \cite{Kim2008PTPS, Gonzalez2022PRD}. By providing more accurate QNM frequencies, our study resolves discrepancies and enhances the reliability of QNM data for Morris-Thorne wormholes. This correction is critical for future endeavours that rely on precise QNM frequencies for theoretical modelling and observational analysis.

Our method employs a Chebyshev-type spectral approach, which allows for efficient and high-precision computation of QNM frequencies. We used 200 polynomials and a precision of 200 digits to ensure robust and accurate results. Such computational efficiency and precision make the spectral approach a powerful tool for studying QNMs. This technique not only improves accuracy but also enables the computation of additional overtones and fundamental frequencies for higher angular momentum modes,  which were not computed in previous studies.

Last but not least, the spectral method we used has been successfully employed by \cite{Batic2024EPJC} to demonstrate the stability of the Noncommutative inspired Schwarzschild black hole proposed by \cite{Nicolini2006PLB} under scalar, electromagnetic, and gravitational perturbations. Its successful application to a different but related problem highlights its robustness and versatility. This provides further validation of the method's effectiveness and reliability in analyzing the stability of exotic spacetime geometries.

Our future endeavours will focus on expanding this study in several key directions. Firstly, we plan to investigate the QNMs in the context of massive perturbations. This extension will allow us to explore the influence of mass on the stability of phantom wormholes, offering not only further insights into their complex nature but also an alternative methodology to that adopted by \cite{Azad2023, Azad2023PRD, Gonzalez2022PRD}. Additionally, we aim to apply our methods to study QNMs of noncommutative geometry-inspired wormholes \cite{Garattini2009PLB, Nicolini2010CQG}. This will broaden our understanding of wormhole physics and provide a comparative framework to evaluate the similarities and differences between black holes and wormholes within the context of noncommutative geometry. Another intriguing direction for future research involves the investigation of the so-called dirty black holes as proposed by \cite{Nicolini2010CQG}. These black holes, characterised by additional matter fields surrounding them, present a more realistic and complex scenario. Analysing their QNMs will shed light on how external matter influences black holes' stability and quasinormal spectra, further enriching our understanding of these objects. In summary, our research opens new avenues for exploring the intricate dynamics of noncommutative geometry-inspired black holes and wormholes. By extending our analysis to include massive perturbations and dirty black holes, we anticipate uncovering new aspects of these entities, contributing to the evolving gravitational and astrophysical research landscape.

\section*{Code availability}

All analytical calculations presented in this document have been verified using the computer algebra system \textsc{Maple}. For transparency and reproducibility, we have included three \textsc{Maple} worksheets that correspond to the analyses conducted in Section~\ref{wholeR}, as well as Subsections~\ref{equivalent} and ~\ref{equivalent1}, within the supplementary materials. The discretization of differential operators \eqref{L0none}--\eqref{L2none} using the Chebyshev-type spectral method is equally performed in \textsc{Maple} computer algebra system. Finally, the numerical resolution of the derived quadratic eigenvalue problem, denoted by equation \eqref{eq:eig}, is executed in the \textsc{Matlab} environment utilizing the \texttt{polyeig} function. Access to all these resources is provided through the following repository link, ensuring that interested parties can freely review and utilize the computational methodologies employed in our study:

\begin{itemize}
    \item \url{https://github.com/dutykh/morristhorne/}
\end{itemize}

\section*{Acknowledgements}

This publication is based upon work supported by the Khalifa University of Science and Technology under Award No. FSU$-2023-014$.

\end{document}